\documentclass[%
aps, 
prl,
 amsmath,amssymb,
reprint,
]{revtex4-1}

\usepackage{graphicx}
\usepackage{dcolumn}
\usepackage{bm}
\usepackage{color}

\usepackage{amsmath}	
\begin{document}

\newcommand{\diff}[2]{\frac{d#1}{d#2}}
\newcommand{\pdiff}[2]{\frac{\partial #1}{\partial #2}}
\newcommand{\fdiff}[2]{\frac{\delta #1}{\delta #2}}
\newcommand{\bx}{\bm{x}}
\newcommand{\bv}{\bm{v}}
\newcommand{\tu}{\tilde{u}}
\newcommand{\br}{\bm{r}}
\newcommand{\bq}{\bm{q}}
\newcommand{\ba}{\bm{a}}
\newcommand{\by}{\bm{y}}
\newcommand{\bY}{\bm{Y}}
\newcommand{\bF}{\bm{F}}
\newcommand{\bn}{\bm{n}}
\newcommand{\be}{\bm{e}}
\newcommand{\new}{\nonumber\\}
\newcommand{\abs}[1]{\left|#1\right|}
\newcommand{\tr}{{\rm Tr}}
\newcommand{\HH}{{\mathcal H}}
\newcommand{\OO}{{\mathcal O}}
\newcommand{\var}{{\rm Var}}
\newcommand{\ave}[1]{\left\langle #1 \right\rangle}

\preprint{AIP/123-QED}

\title{
Minimum scaling model and exact exponents for the Nambu-Goldstone
modes in the Vicsek Model
}
\author{Harukuni Ikeda}
 \email{harukuni.ikeda@gakushuin.ac.jp}
\affiliation{
Department of Physics, Gakushuin University, 1-5-1 Mejiro, Toshima-ku, Tokyo 171-8588, Japan}


\date{\today}



\begin{abstract}
We investigate the scaling behavior of Nambu-Goldstone (NG) modes in the
ordered phase of the Vicsek model, introducing a phenomenological
equation of motion (EOM) incorporating a previously overlooked
non-linear term. This term arises from the interaction between velocity
fields and density fluctuations, leading to new scaling behaviors. We
derive exact scaling exponents in two dimensions, which reproduce the
isotropic scaling behavior reported in a prior numerical simulation.
\end{abstract}

\maketitle

\paragraph{Introduction.---}

When a continuous symmetry is broken spontaneously, there emerge soft
modes known as Nambu-Goldstone (NG)
modes~\cite{nambu1960,goldstone1962}. It is known that continuous
symmetry is not broken in the equilibrium state of two-dimensional
models with short-range interactions since strong fluctuation along the
NG mode destroys the long-range order. This principle is widely
recognized as the Hohenberg-Mermin-Wagner (HMW)
theorem~\cite{hohenberg1967,mermin1966}.

The HMW theorem does not hold out of equilibrium.  Several
out-of-equilibrium systems show continuous symmetry breaking even in
$d=2$~\cite{vicsek1995,xy1995,reichl2010,loos2022long,dadhichi2020,leonardo2023,ikeda2023cor,ikeda2023does,ikeda2024harmonic}.
A prototypical example is the Vicsek model~\cite{vicsek1995}. The model
consists of $XY$ spins moving at a constant speed in their orientation
direction, which mimics the flocking behavior of active matter such as
birds and bacteria~\cite{vicsek1995,nishiguchi2017}.  Numerical
simulations in $d=2$ revealed that the model undergoes a continuous
symmetry breaking from the disordered phase, where the mean velocity of
the spins vanishes, $\ave{\bv}=0$, to the ordered phase, where
$\ave{\bv}\neq 0$~\cite{vicsek1995}.

In 1995, Toner and Tu proposed a hydrodynamic theory to model the
flocking behavior of active matter, such as the Vicsek
model~\cite{toner1995}. Their theory (TT95) successfully explained the
existence of the ordered phase of the Vicsek model in
$d=2$. Furthermore, they derived the exact scaling behavior of the NG
modes in $d=2$, using an elegant argument combining a pseudo-Galilean
transformation and perturbative renormalization group (RG)
methods~\cite{toner1995}. Moreover, a recent more detailed investigation
has shown that their results also become exact in $d>2$, if the system
can be considered as \textit{incompressible}
fluids~\cite{chen2018}. However, for compressible fluids, subsequent
re-analysis of the hydrodynamic theory by Toner, one of the authors of
TT95, pointed out that several additional terms were missing in the
original theory~\cite{toner2012}. These terms hinder the exact
calculations of the scaling exponents even in $d=2$. Consequently, the
applicability of TT95 to compressible fluids, like the Vicsek model,
remains uncertain~\cite{toner2012}.

Numerical studies of the Vicsek model for moderate system size $N\sim
10^6$ reported results consistent with
TT95~\cite{sound1998,giavazzi2017giant}. However, a recent extensive
numerical simulation for larger system size $N\sim 10^9$ uncovered
scaling behaviors that are distinct from the predictions of
TT95. Notably, the simulation indicates almost isotropic scaling in the
ordered phase~\cite{mahault2019}, while TT95 predicts anisotropic
scaling~\cite{toner1995}. Similar isotropic scaling was observed in a
recent experiment of Janus colloidal particles in an AC electric
field~\cite{iwasawa2021}.

Here, we reconcile the above discrepancy between the theory and
numerical simulation. We introduce a simple phenomenological equation of
motion (EOM) for the NG modes that dominate the scaling behavior in the
ordered phase~\cite{nambu1960,goldstone1962}.  A critical insight is
that the symmetry of NG modes differs from the velocity field due to the
hybridization of the velocity field and density fluctuations.
Consequently, our EOM incorporates an additional non-linear term that
was overlooked in previous studies. We calculate the exact scaling
exponents of the EOM in $d=2$. The new scaling exponents obtained by this work
confirm the isotropic scaling observed in the recent numerical
simulation of the Vicsek model~\cite{mahault2019}.

\paragraph{Previous works.---}
Here, we briefly summarize previous theoretical and numerical
studies. We first discuss the \textit{minimal} hydrodynamic theory
introduced and investigated in Ref.~\cite{basse2022}:
\begin{align}
\pdiff{\bm{v}}{t}+\lambda(\bm{v}\cdot\nabla) \bm{v}
= \nabla^2 \bm{v} -
\fdiff{F[\bm{v}]}{\bm{v}}
+\bm{\xi},\label{203525_31Dec23}
\end{align}
where $\bm{\xi}$ denotes the thermal white noise, and $F$ denotes the
standard quartic potential
${F[\bv]=a(\bv\cdot\bv)/2+b(\bv\cdot\bv)^2/4}$ with $b>0$. Here, $a>0$
and $a<0$ represent the disordered and ordered phases,
respectively~\cite{nishimori2011elements}. When the advection $\lambda$
is zero and in $d=2$, Eq.~(\ref{203525_31Dec23}) reduces to the standard
Langevin equation for the XY model and only shows a quasi-long-range
order, which disappears in the thermodynamic
limit~\cite{kosterlitz1972long,kosterlitz2018ordering}. With advection
$\lambda\neq 0$, on the contrary, the model exhibits the true long-range
order~\cite{basse2022}~\footnote{The stability of the ordered phase of
the minimal model remains under debate. A numerical simulation reported
nucleation that destabilizes the ordered phase for moderate
$a$~\cite{basse2022}. However, the characteristic timescale of the
nucleation rapidly increases as $a$ decreases, making it difficult to
discuss whether similar nucleation occurs for small $a$.}. A significant
benefit of the minimal model (\ref{203525_31Dec23}) is that the analysis
in TT95 becomes exact for this model. Numerical results of
(\ref{203525_31Dec23}) in the ordered phase indeed well agree with
TT95~\cite{basse2022}.

Compared to the full hydrodynamic theory by Toner and
Tu~\cite{toner1995}, the minimal model (\ref{203525_31Dec23}) neglects
the effects of the density fluctuations $\delta\rho$, which appear
through the pressure-gradient forces such as $\nabla P$ and
$\bv(\bv\cdot\nabla)P$.  Here, the pressure $P$ depends on both
$\delta\rho$ and $\bv$.  Due to this complexity, the RG calculations of the full hydrodynamic theory are extremely
challenging, and exact scaling exponents are yet to be known even in
$d=2$~\cite{toner2012}. In this work, instead, we propose a simple
phenomenological EOM for the NG modes, which can be treated exactly in
$d=2$, as we will see below.

\paragraph{Decoupling of NG modes.---}

In the ordered phase of the Vicsek model, the NG modes are hybridization
between velocity fluctuations perpendicular to the order parameter
$\delta\bv_\perp$ and density fluctuations $\delta\rho$.  Here, we
discuss that those NG modes are decoupled into non-interacting
eigenmodes in the long-time limit.

The EOM of the slow modes can be obtained by using the gradient
expansion~\cite{toner1998,toner2012}. The analysis can be simplified in
$d=2$ because the perpendicular velocity has only one component $\delta
v_\perp$. The gradient expansion of the EOM can be generally written as
\begin{align} 
\begin{pmatrix}
\delta\dot{v}_\perp \\ \delta\dot{\rho} 
\end{pmatrix}
 = 
\begin{pmatrix}
\bm{A} & \bm{B}\\
\bm{C} & \bm{D}
\end{pmatrix}
\begin{pmatrix}
\nabla\delta v_\perp \\ \nabla \delta\rho 
\end{pmatrix} + \cdots\label{203329_12Oct24} 
\end{align}
where $\bm{A}$, $\bm{B}$, $\bm{C}$, and $\bm{D}$ denote unknown 
vectors. In the Fourier space $f(\bq,t)\equiv \int d\bx
f(\bx,t)e^{i\bq\cdot\bx}$, the EOM can be diagonalized, leading to the
eigenmodes $\{\tilde{u}_\mu(\bq,t)\}_{\mu=\pm}$. For small wave number
$\bq$, the EOM of the eigenmodes can be written as 
\begin{align}
\dot{\tilde{u}}_\mu = iq\Omega_\mu(\hat{q})\tilde{u}_\mu 
+ O(q^2, \tilde{u}_\mu^2),\label{114730_10Jun24}
\end{align}
where $q=\abs{\bq}$, $\hat{q}=\bq/q$,
and the sound velocities are 
\begin{align}
\Omega_\pm(\hat{q}) = \frac{A_q+D_q \pm \sqrt{(A_q^2-D_q^2)^2+4B_qC_q}}{2}.
\end{align}
Here, $X_q=\bm{X}\cdot\hat{q}$ denotes the projection of a vector
$\bm{X}$ onto the direction $\hat{q}$. The linear advection term
$iq\Omega_\mu\tilde{u}_\mu$ in (\ref{114730_10Jun24}) can be eliminated
by using the interaction representation:
\begin{align}
u_\mu(\bq,t)=\tu_\mu(\bq,t)e^{-i\Omega_\mu(\hat{q})t}.\label{165833_8Jun24}
\end{align}
It is crucial to note that the eigenmodes with different sound
velocities are uncorrelated~\cite{toner1998,chate2024dynamic}. This can
be seen by observing the equal-time correlation averaged over the total
volume $V$~\cite{hansen2013theory}:
\begin{align}
&\ave{\tu_\mu^*(\bx,t)\tu_\nu(\bx',t)}
 \equiv \frac{1}{V}\int d\bx'' \tu_\mu^*(\bx+\bx'',t)\tu_\nu(\bx'+\bx'',t)\new 
& =\frac{1}{V(2\pi)^2}\int d\bq u_\mu^*(\bq,t)u_\nu(\bq,t)e^{i\bm{q}\cdot (\bx-\bx')}
 e^{iq(\Omega_\mu(\hat{q})-\Omega_\nu(\hat{q}))t},
 \label{172621_9Jun24}
\end{align}
which gives a negligible contribution for $\mu\neq \nu$, due to the fast
oscillation of the factor $e^{iq(\Omega_\mu-\Omega_\nu)t}$ for $t\gg
1$. This enables to treat the eigenmodes $u_\pm$, separately, see the
full paper of TT95~\cite{toner1998} and the Supplementary information of
Ref.~\cite{chate2024dynamic} for complementary and more rigorous
discussions. Note that $\delta v_\perp$ and $\delta\rho$ are linear
combinations of the eigenmodes $u_\pm$, implying that they have the same
scaling as $u_\pm$, $\delta v_\perp\sim \delta \rho\sim u_{\pm}$.

The same argument also holds in $d=3$, since the
diagonalization of the EOM~(\ref{203329_12Oct24}) in $d=3$ leads to
three eigenmodes propagating with different sound
velocities~\cite{toner1998}.

\paragraph{New phenomenological EOM.---}
Since $\{u_{\mu}(\bx,t)\}_{\mu=1,\cdots, d}$ do not interact, we can
construct and investigate their EOM separately.  To simplify the
notation, hereafter, we omit the subscript $\mu$ of $u_\mu$, unless
otherwise specified.
The most general form of the EOM of $u$ is given by
\begin{align}
\dot{u} = \nabla^2u + f_{\rm nl}[u] + \xi,\label{155425_17Jun24}
\end{align}
where $f_{\rm nl}$ denotes (unknown) non-linear
terms, and $\xi$ denotes the white noise of zero mean and
variance
\begin{align}
\ave{\xi(\bx,t)\xi(\bx',t')}
=2T\delta(\bx-\bx')\delta(t-t').\label{180337_13Feb24}
\end{align}

Let us first consider the linear model with $f_{\rm nl}=0$. To investigate
the large spatio-temporal scaling behavior, we apply the following
scaling transformations~\cite{burger1989,toner1995}:
\begin{align}
\bx_\perp\to l\bx_\perp,\ x_\parallel \to l^{\zeta}x_\parallel,\
t\to l^{z}t,\
u\to l^{\chi}u,
\end{align}
where and hereafter the subscripts $\parallel$ and $\perp$ denote
components of any vector parallel and perpendicular to the mean velocity
$\ave{\bm{v}}$, respectively.  Using (\ref{180337_13Feb24}), we find
the scaling dimension of the noise as $\xi\to
l^{-(z+\zeta+d-1)/2}\xi$~\cite{burger1989}. Then, the EOM
reduces to
\begin{align}
l^{\chi-z}\dot{u} = l^{\chi-2}\nabla_\perp^2 u
 + l^{\chi-2\zeta}\partial_\parallel^2 u + l^{-(z+\zeta+d-1)/2}\xi.
\end{align}
Requiring that the all terms have the same scaling dimension, we get the
following scaling relations:
\begin{align}
&\chi-z = \chi-2\\
&\chi-z = \chi-2\zeta\label{204731_13Feb24}\\
&\chi-z = -\frac{z+\zeta+d-1}{2},\label{204745_13Feb24}
\end{align}
leading to~\cite{burger1989,toner1995}
\begin{align}
z = 2,\ \zeta = 1,\ \chi = \frac{2-d}{2}. 
\end{align}
For $d<2$, the fluctuation $\ave{\delta u^2}\sim l^{2\chi}$ diverges in
the limit $l\to\infty$, implying that the fluctuations of the NG modes
destroy the long-range order. Therefore, the lower critical dimension
for continuous symmetry breaking is $d_{\rm low}=2$. The result is
consistent with the HMW
theorem~\cite{hohenberg1967,mermin1966,tasaki2020}.

Next, we investigate the effects of the non-linear terms. The analysis
can be simplified in $d=2$, where $\nabla_\perp \to
\partial_\perp$. Above the lower critical dimension, $u\sim l^{\chi}$,
$\partial_\perp\sim l^{-1}$, and $\partial_\parallel\sim l^{-\zeta}$
vanish in the thermodynamic limit $l\to\infty$, enabling the expansion
of the non-linear term $f_{\rm nl}$ by $u$ and its spatial derivatives,
$\partial_{\perp{\rm or}\parallel}u$, $\partial_{\perp{\rm
or}\parallel}^2u$, $\cdots$. Since the restoring force of the NG mode
should vanish for spatially uniform deformations, each term in $f_{\rm
nl}$ should involve at least one spatial derivative $\partial_{\perp{\rm
or}\parallel}$~\cite{altland2010condensed}. Therefore, $f_{\rm nl}$ can
be expanded as
\begin{align}
f_{\rm nl}[u,\partial_{\perp{\rm or}\parallel} u,\partial_{\perp{\rm or}\parallel}^2u,\cdots]&
= c_1u\partial_\perp u + c_2 u\partial_\parallel u\new 
&\hspace{-1cm}+ O\left(u^2\partial_{\perp{\rm or}\parallel} u, u(\partial_{\perp {\rm or}\parallel} u)^2,u\partial_{\perp {\rm or}\parallel}^2 u\right),
\end{align}
where $c_1$ and $c_2$ are constants. The higher-order terms have smaller
scaling dimensions and are irrelevant for ${l\gg 1}$. In summary, the
EOM in $d=2$ reduces to
\begin{align}
\dot{u} &\approx 
(\partial_\perp^2+\partial_\parallel^2)u
 +c_1 u\partial_\perp u+ c_2 u\partial_\parallel u +\xi.
 \label{114525_12Feb24}
\end{align}
Below, we shall argue that the non-linear terms stabilize the long-range
order and enable the continuous symmetry breaking in $d=2$.

\paragraph{Exact exponents in two dimensions.---}

For the minimal hydrodynamic theory (\ref{203525_31Dec23}), the NG mode
is identified with the velocity component orthogonal to the order
parameter $u=v_\perp$. In this case, the EOM must remain invariant under
the transformation $x_\perp \to -x_\perp$ and $v_\perp\to -v_\perp$,
necessitating $c_2=0$~\cite{sartori2019}. This specific case,
(\ref{114525_12Feb24}) with $c_2=0$, has been well investigated in
previous studies~\cite{toner1995,sartori2019}. For completeness, we will
briefly outline the derivation of the scaling exponents. First, note
that the non-linear term can be rewritten as a total derivative with
respect to $\partial_\perp$ as $c_1 u \partial_\perp u = \partial_\perp
(c_1 u^2/2)$. A perturbative calculation of this term generates
terms such as $\partial_\perp (\text{some function})$, which only affect
the scaling dimensions of terms involving $\partial_\perp$. Therefore,
the scaling dimensions $\dot{u}\sim l^{\chi-z}$, $\partial_\parallel^2
u\sim l^{\chi-2\zeta}$, and $\xi\sim l^{-(z+\zeta+d-1)/2}$ remain
unchanged. Requiring that those three terms have the same scaling
dimension, we get the scaling relations (\ref{204731_13Feb24}), and
(\ref{204745_13Feb24}) for $d=2$. Next, note that (\ref{114525_12Feb24})
with $c_2=0$ is invariant under the pseudo-Galilean transformation: $u
\to u + U$, $x_\perp \to x_\perp + c_1 Ut$, which implies the following
scaling relation~\cite{toner1995,toner2005}:
\begin{align}
1 = \chi + z.\label{205404_13Feb24} 
\end{align}
Using (\ref{204731_13Feb24}), (\ref{204745_13Feb24}), and (\ref{205404_13Feb24}),
we get~\cite{sartori2019}
\begin{align}
z = \frac{2(1+d)}{5},\ \zeta = \frac{d+1}{5},\ \chi = \frac{3-2d}{5},\label{205755_13Feb24}
\end{align}
which are consistent with TT95~\cite{toner1995}. Note that the scaling
relations have been proven only in $d=2$, and it is not clear if the
exponents (\ref{205755_13Feb24}) become exact in $d>2$.  Remarkably, a
recent detailed investigation has established that the exponents
(\ref{205755_13Feb24}) become exact for incompressible fluids
$\nabla\cdot\bv=0$ in $d>2$~\cite{chen2018}. The scaling behaviors
(\ref{205755_13Feb24}) imply $\ave{\delta v_\perp^2}\sim l^{2\chi}\to 0$
in $d=2$, meaning that the non-linear term stabilizes the long-range
order in $d=2$~\cite{toner1995}. The exponents (\ref{205755_13Feb24})
well agree with a numerical simulation of the minimal model
(\ref{203525_31Dec23}) in $d=2$~\cite{basse2022}.

Let us turn to the Vicsek model. It is crucial to note that the velocity
field generally couples to the density fluctuations~\cite{toner2012},
and hence, the NG mode is not identical to $v_\perp$. In other words,
there is no reason to assume $c_2=0$. When constructing a
phenomenological equation, all the relevant terms allowed by the
symmetry must be taken into account.  Below, we calculate the scaling
exponents for the case $c_2\neq 0$. In (\ref{114525_12Feb24}), the
non-linear terms proportional to $c_i$ can be rewritten as the total
derivatives with respect to $\partial_\parallel$ and $\partial_\perp$,
meaning that the perturbative RG calculations yield
no corrections to the scaling of $\dot{u}$ and
$\xi$~\cite{toner1995,mahault2019,sartori2019}. Requiring $\dot{u}\sim
\xi$, we get (\ref{204745_13Feb24}) for $d=2$~\cite{mahault2019}.  This
scaling relation is often referred to as the hyperscaling and has been
confirmed in the recent numerical simulation of the Vicsek
model~\cite{mahault2019}. Note also that (\ref{114525_12Feb24}) is
invariant under the pseudo-Galilean transformation: $u\to u + U$,
$x_\perp \to x_\perp + c_1 Ut$, and $x_\parallel\to x_\parallel + c_2
Ut$, which implies the following scaling
relations~\cite{toner1995,toner2005}
\begin{align}
&1= \chi + z,\label{142914_18Jan24} \\
&\zeta= \chi + z.\label{164004_13Feb24}
\end{align}
Using (\ref{204745_13Feb24}), (\ref{142914_18Jan24}), and
(\ref{164004_13Feb24}),
we can determine the scaling exponents:
\begin{align}
z = \frac{d+2}{3},\ \zeta = 1,\ \chi = \frac{1-d}{3}.\label{111507_14Feb24}
\end{align}
The results are consistent with our previous heuristic scaling
argument~\cite{ikeda2024advection}. Given that (\ref{114525_12Feb24}) is
the most general form of EOM taking into account the lowest order
non-linear terms, we conclude that (\ref{111507_14Feb24}) are the exact
scaling exponents of the Vicsek model in $d=2$. In $d>2$, it is unclear
if the EOM satisfies the scaling relations (\ref{204745_13Feb24}),
(\ref{142914_18Jan24}), and (\ref{164004_13Feb24}), and we can not rule
out the possibility that some corrections appear to the exponents
(\ref{111507_14Feb24}).

The exponents~(\ref{111507_14Feb24}) in $d=2$ are exact within the
framework of perturbative RG. However, strictly speaking, we cannot rule
out the possibility of non-perturbative
contributions~\cite{canet2010,canet2011,dupuis2021nonperturbative}. We
next compare our results with numerical simulations to confirm that,
even if such contributions exist, they are very small.

\paragraph{Comparison with numerical simulation.---}

\begin{table}[t]
\begin{center}
\caption{Scaling exponents. Data are taken from
Ref~\cite{mahault2019}.}  \label{235502_31Dec23}
\begin{tabular}{l|rrr|rrr}
&$d = 2$ &  & & $d=3$ & & \\ 
& Vicsek& TT95& This work & Vicsek&TT95& This work \\ \hline
$\chi$ & -0.31(2)& -0.2 & -0.33& -0.62 &-0.6 &  -0.67\\
$\zeta$ & 0.95(2) & 0.6 &1 &  1 &0.8  &1 \\
$z$ &1.33(2)& 1.2&  1.33 & 1.77 & 1.6&  1.67 \\
$\alpha$ &0.84(1)& 0.8&  0.83& 0.79(2) & 0.77&  0.78
\end{tabular}
\end{center}
\end{table}
In Table~\ref{235502_31Dec23}, we compare the numerical results of the
Vicsek model~\cite{mahault2019}, TT95~\cite{toner1995}, and our
theoretical predictions (\ref{111507_14Feb24}). Overall, our
results show better agreement with numerical results than that of TT95.
In particular, we get an almost perfect agreement for $d=2$. For $d=3$,
there are small discrepancies between the numerical and theoretical
results. We expect that further
theoretical~\cite{toner1995,toner2005,chen2016mapping,sartori2019,diCarlo2022,jentsch2023},
and numerical studies~\cite{gregoire2004,mahault2019} will lead to more
accurate estimates of the scaling exponents in $d=3$.

\begin{figure}[t]
\includegraphics[width=8cm]{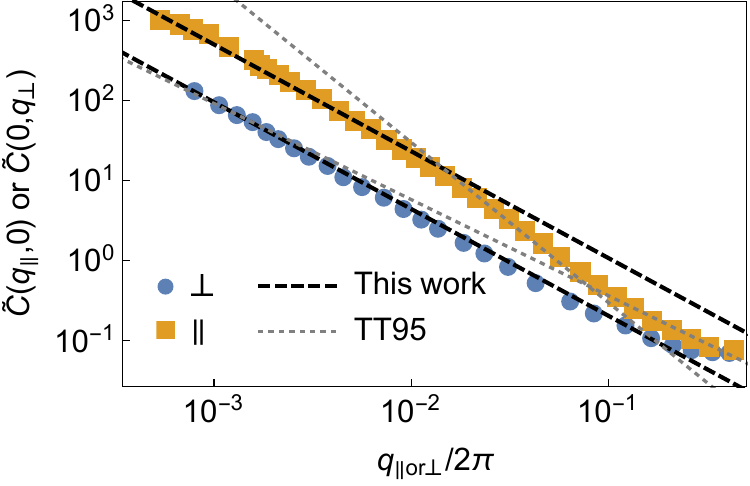} \caption{Correlation
 functions. Markers denote the numerical results of the Vicsek model in
 $d=2$ for $L=8000$, taken from Fig.1~(a) in
 Ref.~\cite{mahault2019}. The dashed and dotted lines represent
 theoretical predictions.}  \label{083952_8Feb24}
\begin{center}
\end{center}
\end{figure}

Another important physical quantity is the correlation function
$C(\bx)=\ave{\bv_\perp(\bx)\cdot \bv_\perp(0)}$. The scaling behaviors
$\bx_\perp\sim l^{}$, $x_\parallel\sim l^\zeta$, and $\bv_\perp \sim
u\sim l^{\chi}$ lead to
\begin{align}
C(\bx) \sim l^{2\chi}C(l^{-1}\bx_\perp,l^{-\zeta}x_\parallel)
 \sim \abs{\bx_\perp}^{2\chi}\sim x_\parallel^{2\chi/\zeta}.
\end{align}
In the Fourier space, it leads to~\cite{toner1995,toner1998,toner2005}:
\begin{align}
\tilde{C}(\bm{q})
&=\int d\bx e^{i\bm{q}\cdot\bx}C(\bx) \new
&\sim l^{2\chi+\zeta+d-1}\tilde{C}(l\bm{q}_\perp,l^\zeta q_\parallel)
 \sim \abs{\bm{q}_\perp}^{-z}\sim q_\parallel^{-z/\zeta}.
\end{align}
To precisely calculate $\tilde{C}(\bm{q})$, one should perform a numerical
simulation for a sufficiently large linear system size $L$ because the
minimal wave vector $q_*$ scales as $q_*\sim 2\pi/L$. The best
numerical result for the Vicsek model has been obtained in $d=2$ for
$L=8000$ in Ref.~\cite{mahault2019}. In Fig.~\ref{083952_8Feb24}, we
compare the numerical result, our theoretical prediction, and that of
TT95~\cite{toner1995,toner1998}. Our theory shows a better agreement
with the numerical results than TT95.

In the ordered phase of the Vicsek model, the density fluctuations are
significantly enhanced due to the NG modes. This phenomenon is referred
to as the giant number fluctuations (GNF)~\cite{nishiguchi2023}. The GNF
is popularly studied in numerical and experimental
studies~\cite{mahault2019,iwasawa2021,nishiguchi2023}. So, we here
derive the scaling exponent $\alpha$ that characterizes the strength of
the GNF. To be more precise, we shall observe the number fluctuations in
a box of linear size $l$:
\begin{align}
\ave{\delta N^2} &= \ave{\left(\int_{[0,l]^d}d\bx\delta\rho(\bx)\right)^2}\new 
&= l^d \int_{[0,l]^d} d\bx \ave{\delta\rho(\bx)\delta\rho(0)}.\label{150350_15Feb24}
\end{align}
We assume that the density fluctuations are dominated by the NG
modes, $\delta\rho\sim u$, and the correlation function
$C_\rho(\bx)=\ave{\delta\rho(\bx)\delta\rho(0)}$ satisfies 
$C_\rho(\bx_\perp,x_\parallel)=l^{2\chi}
C_\rho(l^{-1}\bx_\perp,l^{-\zeta}x_\parallel)$. Then, the scaling analysis 
leads to $\int d\bx C_\rho(\bx)\sim l^{2\chi+\zeta+d-1}\sim l^z$.
Substituting this into (\ref{150350_15Feb24}), we get
\begin{align}
\ave{\delta N^2} \sim l^{d+z} \sim (N/\rho)^{2\alpha},
\end{align}
where $\rho=N/l^d$ denotes the mean-density, and we have introduced the
exponent~\cite{toner1995,toner1998}
\begin{align}
\alpha = \frac{1}{2} + \frac{z}{2d}.\label{155423_15Feb24}
\end{align}
The central limit theorem predicts $\alpha=1/2$. In the ordered phase of
the Vicsek model, one has $\alpha>1/2$: the fluctuations grow much
faster than the naive expectation of the central limit
theorem~\cite{nishiguchi2023}. This anomalous enhancement of the density
fluctuations is the signature of the GNF~\cite{nishiguchi2023}. In
Table~\ref{235502_31Dec23}, we compare the numerical
result~\cite{mahault2019,nishiguchi2023}, our theoretical prediction,
and that of TT95~\cite{toner1995}. In $d=2$, our theory shows a better
agreement with the numerical result than that of TT95. In $d=3$, the
predictions of the two theories are so close that we can not judge which
theory is more precise from the current numerical data. Extensive
numerical simulations for larger system sizes are required to elucidate
this point.

\paragraph{Summary and discussions.---}
In summary, we have investigated the NG modes of the Vicsek model in the
ordered phase. We derived a phenomenological EOM of the NG modes and
investigated its scaling behavior. Our essential observation is that the
NG modes should not be simply identified with the vertical components of
the velocity field but should depend both on the velocity field and the
density fluctuations. As a consequence, the EOM has an additional
non-linear term overlooked in the previous studies, which leads to a new
scaling behavior. We calculated the exponents that characterize the
scaling of the NG modes. In particular, the pseudo-Galilean invariance
allows us to calculate the exact scaling exponents in $d=2$. We found
that our results agree well with a recent numerical simulation of the
Vicsek model in $d=2$.

Although we derived the EOM by a phenomenological argument, the
resultant EOM incorporates all the relevant non-linear terms, making it
quite universal. Therefore, we expect that the EOM can be applied to
other two-dimensional systems that exhibit long-range order, as well as
the Vicsek model. For instance, two-dimensional long-range orders have
been reported by recent experiments with bacteria and Janus
particles~\cite{nishiguchi2017,iwasawa2021}. It is an important future
work to check if these systems follow the scaling laws found in this
work.

\paragraph{Note added.---}

While making final revisions, two relevant papers
have come to our attention. 
Jentsch and Lee addressed the same problem of the scaling exponents of
the Vicsek model~\cite{jentsch2024new}.  They introduced and
investigated a simplified version of the Toner-Tu hydrodynamic
theory~\cite{toner1995,toner1998,toner2012}. Their RG calculations for
the simplified hydrodynamic theory predict weak anisotropy:
$\zeta=0.975$ in $d=2$. It remains an intriguing question whether
similar calculations for the full hydrodynamic theory can reproduce the
isotropic scaling behavior $\zeta=1$.
Chat{\'e} and Solon also addressed the scaling exponents of the polar
flocks in $d=2$~\cite{chate2024dynamic}. They pointed out that a
previously overlooked cubic term $\partial_\parallel u^3$ is actually
relevant for the minimal model~(\ref{203525_31Dec23}), leading to
different scaling exponents from TT95~(\ref{205755_13Feb24}). This,
however, does not change the scaling exponents of the Vicsek
model~(\ref{111507_14Feb24}) because it is irrelevant compared to
$u\partial_\parallel u$~\cite{chate2024dynamic}.

\acknowledgments 
We thank H.~Tasaki for his careful reading of the manuscript and
constructive comments. We also thank H.~Nakano, Y.~Kuroda,
D.~Nishiguchi, H.~Chat{\'e}, and A.~Solon for helpful discussions. This
work was supported by KAKENHI 23K13031.



\bibliography{reference}

\end{document}